\documentclass[aps,showpacs,nofootinbib]{revtex4}
\usepackage{epsf,latexsym}

\begin{document}

\title{Entropy bounds and field equations}
\author{Alessandro Pesci}
\email{pesci@bo.infn.it}
\affiliation
{INFN-Bologna, Via Irnerio 46, I-40126 Bologna, Italy}

\begin{abstract}
For general metric theories of gravity,
we
compare
the approach 
that
describes-derives the field equations
of gravity as a thermodynamic identity 
with the one which looks at them
from entropy bounds.
The comparison is made 
through 
the
consideration
of the matter entropy flux across (Rindler) horizons,
studied
by
making use of the notion of 
a
limiting thermodynamic scale $l^*$ 
of matter,
previously introduced
in the context of entropy bounds.

In doing this: 
i) a bound 
for
the entropy of any lump of matter
with 
a given
energy-momentum tensor $T_{ab}$
is considered, 
in terms of a quantity 
which is 
independent of the theory of gravity we use;
this quantity is the variation of 
the
Clausius entropy
of a suitable horizon when the element of matter crosses it;
ii) 
by
making use of the equations of motion of the theory,
the same quantity is then expressed 
as 
the
variation of Wald's entropy of that horizon
(and this leads to
a generalized form of the generalized covariant entropy bound, 
applicable
to general 
diffeomorphism-invariant
theories
of gravity);
iii) a notion of $l^*$ for horizons,
and an expression for it, is given.

\end{abstract}


\maketitle

$ $
\section{Motivation}

The thermodynamic description of the field equations of gravity
was
first performed by Jacobson in \cite{JacA},
where these equations 
were
shown to be 
equivalent to
a thermodynamic identity.
In \cite{PadE, PadF},
Padmanabhan
has shown 
that this identity was obtainable from the extremisation
of a suitable entropy functional
(see also \cite{PadA, PadB} for a general account 
of
this).
In these approaches,
arbitrary accelerating observers 
at a generic point of spacetime
are considered.
The equations of motion are obtained through 
the
proper consideration
of a key quantity, 
$\frac{dE}{T_H}$
(energy flux of matter crossing the horizon
over horizon temperature), 
or variation of the Clausius entropy of the horizon,
associated 
with
the crossing of matter 
through local Rindler horizons.
It represents the
change of entropy
of the ``system beyond the horizon''
or, in brief, of the horizon
(concerning the general issue 
of whether
patches of accelerating
(or general) horizons,
i.e. not only of black hole horizons, can consistently
be associated 
with
an entropy, cf. \cite{JacC}).

Our aim here,
is to contrast this approach with the one which 
obtains
the equations of motion from entropy bounds.
Crucial to these is
the entropy $dS$ of matter. 
In entropy bounds
(in their generalised formulation \cite{Fla}
which can be thought of as subsuming many other formulations
(including the first entropy bound ever formulated,
the Bekenstein bound 
for the
entropy-to-energy ratio \cite{BekD})
and can be considered as a general statement
of holography),
the entropy $dS$ of matter which goes through a null hypersurface
is bounded 
by a quarter of the 
change
of the area
of the spatial cross-section of the hypersurface.
The existence of such a bound
is seen as remounting
at the end
to the existence 
of a lower limit 
to 
the
``amount of space''
to be assigned to the single bit of information,
a limit
of the order of 
the Planck length.

If one further postulates
this latter feature of information
is intrinsic,
that is
unrelated to gravity,
it is then possible to {\it derive}
the equations of motion of gravity from it.
That is, 
from the mere existence of this fundamental discretisation
of 
space
associated 
with
information,
one can derive
that matter curves spacetime, and 
the way it does
(independent explicit implementations of this idea
are \cite{JacB}, \cite{Ver}; 
see also \cite{PesG,PesH};
it can however be 
seen that it was
already contained in \cite{JacA}; 
in \cite{PadD, PadC} and \cite{PadB}, 
a gravitational
acceleration of entropic origin is also considered,
but, apparently, with no intention 
of considering
this as 
a {\it derivation} of
gravity from horizon entropy).
If on the contrary,
on
defining the limiting ``amount of space'' above,
gravity itself 
intervenes,
the {\it derivation}
would turn into a {\it description} of gravity in terms of entropy bounds.
Gravity would be one among the basic players
and would no longer be reducible, at a fundamental level,
to 
the pure
thermodynamics of something else
(while maintaining
its emergent thermodynamic nature 
in the long wavelength limit).

What is behind the formulation of
the entropy bounds
in terms of 
the area/4
of suitable surfaces, 
is that,
in case these surfaces act as 
(spatial cross sections of) horizons,
the area/4 does coincide with
horizon entropy,
and this 
implies
Einstein gravity. 
The approach to the equations of motion as
a thermodynamic identity
is independent, instead,
of being 
the horizon entropy 
given by area times $\frac{1}{4}$ 
or by area times any other constant,
or, also, given by general expressions 
involving, say, 
a dependence on curvature at each point
(we are insisting, pedantically perhaps, on the expression
``area times a constant'' in place of ``proportional to area''
because the latter expression
can include entropies
still locally proportional to area, 
but not given by area times a constant,
due to a dependence on the point).
The results in \cite{JacA} have been obtained, indeed,
on assuming
horizon entropy as area times a constant;
but extensions of these results to
more general expressions for horizon entropy
have been suggested in \cite{JacA} itself,
and explicitly worked out
in \cite{Cai}
for Gauss-Bonnet and 
general Lanczos-Lovelock models \cite{LanA, LanB, Lov}
(in a cosmological setting),
in \cite{Eli}
(for $f(R)$ theories) 
and
(to include any diff-invariant theory)
in \cite{BruC} and \cite{PadA, PadB},
through 
the use
of Wald's entropy \cite{WalB, JacD, Iye} 
as horizon entropy.
The mentioned derivation in \cite{PadE, PadF} 
of the equations of motion
from extremisation of a suitable entropy functional,
lives 
moreover
in a context far more general than Einstein gravity
(Lanczos-Lovelock models). 

In our bridging 
of the
two approaches,
there is thus a need
to disantangle in the entropy bounds
what is gravitational-dynamics dependent
and what can act as
a bound to matter entropy
in the form of a quantity which is irrespective of the gravitational theory
we are in.
This is what we try to do in 
what follows.
 
The framework in which we move
are arbitrary diffeomorphism invariant theories
of gravity with gravitational Lagrangians depending 
on the metric, on the Riemann tensor, and on the derivatives of the
latter of whichever order and nothing else,
and with minimal coupling between matter and gravity.


\section{Bound to the entropy of an element of matter as 
variation of Clausius entropy of the hottest horizon going to swallow it}
\label{section2}

In the calculation of matter fluxes through the horizon,
approaches differ in the assumptions they make
concerning the local Rindler horizon they are considering
(for instance, Jacobson \cite{JacA} and
Padmanabhan \cite{PadE, PadF, PadA, PadB}).
In particular,
the expansion of the null congruence which generates the horizon
is assumed to be vanishing at the point under
consideration in \cite{JacA}, 
while it is arbitrary in \cite{PadE, PadF, PadA, PadB}.
This difference in the assumptions, has remarkable
consequences concerning 
the definition/interpretation of some thermodynamic potentials
associated with
matter fluxes 
through horizons, 
as pointed out and discussed in \cite{KotC}.
%
However it does not appear to affect 
the variation of Clausius
entropy of the horizon,
provided that
a notion of Clausius entropy $\frac{dE}{T_H}$ 
for 
arbitrary bifurcate null surfaces is introduced \cite{Bac}.

These topics
are strictly connected with
present discussion.
Here, however,
we wish
to elaborate on Clausius entropy of the horizon
in its ability to bound the entropy of matter dropped through the horizon
and in its 
independence of the
gravity theory we are in,
regardless of any 
given
horizon expansion.
We choose thus to consider the simplest possible case,
namely the case of a patch of Rindler horizon
with vanishing expansion $\theta$.
The results we 
shall
find are applicable
to any horizon locally appoximable by Rindler $\theta =0$
(as bifurcate Killing horizons are).

The Wald entropy 
associated with
any patch of horizon, 
in particular to Rindler horizons with $\theta =0$,
does depend, of course, on the gravity theory we are in.
The emphasis in this Section 
is that
the variation of Clausius entropy of the horizon
associated to the passage
of a lump of matter through it,
is instead a (gravitational) dynamics-independent quantity.
The reason 
for
this emphasis is 
that,
given any element of matter
characterized
by its energy-momentum tensor $T_{ab}$ besides its size,
our aim is to construct 
a bound to its entropy
in terms of 
the variation
of 
the Clausius entropy
of (a certain) horizon; 
a bound which thus
turns out to be independent of gravitational dynamics.
%
We do this 
by making use of the $l^*$ concept,
introduced earlier \cite{PesB, PesC}.

%
We consider some smooth distribution of matter.
We associate with it 
its energy-momentum tensor field $T_{ab}$
defined
in a $D$-dimensional (gauge \cite{RovC}) spacetime ($D\geq4$)
with metric $g_{ab}$,
the geometry of this spacetime being determined,
through some field equations,
by that matter and all other matter in the universe.
%
At a generic point $P$ in this spacetime, 
let $k^a$ be a null vector, directed to the future of $P$.
We 
then consider 
a local inertial frame around $P$, 
with coordinates $X^a$,
such that $P$ is at $X^a = 0$,
and chosen in such a way 
as to have
$k^a$ given by (1,1)
in the plane
$(T, X)$.
The null curve $(X, X)$
has affine parameter $X$ and tangent $k^a$.
%
We consider a small piece of matter around $P$.
The element of matter has a generic velocity
in the local inertial frame.
Its flow across any patch of surface
is determined by the velocity vector,
which has an intrinsic meaning 
independent of the gravity theory which determined
a given spacetime as solution
(and back-reaction effects, due to shrinking of the area 
while the element of matter goes through, are second order effects
with respect to this in the flow). 

Let us 
next consider
the local Rindler frame \cite{JacA,  PadA}
associated with
an observer accelerating along 
$X$ with acceleration $\kappa$, 
which is
at rest with respect to the local inertial frame at Rindler time $t = 0$.
We know that the metric can be 
written as
$
ds^2 = 
-(kx)^2 dt^2 + dx^2 +dx_{\perp}^2 \equiv
- N^2 dt^2 + dx^2 + dx_{\perp}^2   
$ 
with obvious notation,
with
$x^a$ Rindler coordinates
and $dx_{\perp}^2$ the Euclidean metric
in the $(D-2)$-plane.
In these coordinates,
the accelerating observer $\cal A$ is, at $t = 0$,
at $x = \frac{1}{\kappa}$.

The Killing vector 
$
\xi^{a}_{(t)} \equiv 
\big( \frac{\partial}{\partial t}  \big)^a, 
$
corresponding to time translation invariance,
null on the horizon and orthogonal to it,
is what is needed to express the mass $M$ of the horizon
(of the ``system beyond the horizon''),
attributed to it by the accelerating observer.
The condition
$
\xi_{(t)}^a \xi_{(t)a} = - N^2 = 0,
$
that is $x = 0$, or $T \pm X =0$,
gives the location
of the horizon.
To this,
the accelerating observer, 
which 
sees
herself in a thermal bath,
assigns a temperature 
$T_H = \frac{\kappa}{2\pi}$. 
For the entire construction to be consistent,
the size $L_R \approx \frac{1}{\kappa}$ 
associated to the Rindler description
must be much smaller than the size $L_I$ 
on which the local inertial frame
approximation works, 
that is 
$L_R \ll L_I \approx \frac{1}{\sqrt{\cal R}}$,
where $\cal R$ is the magnitude of a typical component
of the curvature tensor.
We assume to have chosen $\kappa$ large enough 
such that
this is the case.

Let us first review
how it goes in the standard approach,
that the variation of the 
Clausius entropy of the horizon
associated 
with
the passage of an element of matter
is independent of the gravitational dynamics.
%
For this,
we have to 
obtain
the energy flux $dE$ entering
in the
expression $\frac{dE}{T_H}$. 
It coincides
(since the horizon is orthogonal to $\xi^a_{(t)}$)
with the increment in mass $dM$
of the horizon.
Standard calculation gives 

\begin{eqnarray}
dE =
T_{ab} \ \xi^a_{(t)} k^b \ A \ dX,   
\end{eqnarray}
where $A$ is the area of the cross-section
of the element of matter in the $(D-2)$-plane,
and $dX$ its size in $X$-coordinate. 
This quantity depends on the characteristics
of the lump of matter ($T_{ab}$, $A$, $dX$)
and on the geometric characteristics of the horizon
($\xi^a_{(t)}$, $k^b$);
in no way 
can it depend
on the equations of motion of gravity,
whatever they are,
having
$g_{ab}$ as solution 
associated 
with
the assigned 
distribution of matter.   
The same can be said of the associated 
Clausius entropy variation of the horizon,
since 

\begin{eqnarray}\label{Clausius}
\frac{dE}{T_H} =
\frac{1}{T_H} \ T_{ab} \ \xi^a_{(t)} k^b \ A \ dX,
\end{eqnarray}
and $T_H$ is given with the field $\xi^a_{(t)}$.
Thus, the expression
of the amount of Clausius entropy of the horizon
associated 
with the
passage of a lump of matter through it,
displays in an obvious way
the independence of this quantity 
with respect to
gravitational dynamics.

\begin{figure}\leavevmode
\begin{center}
\epsfxsize=8cm
\epsfbox{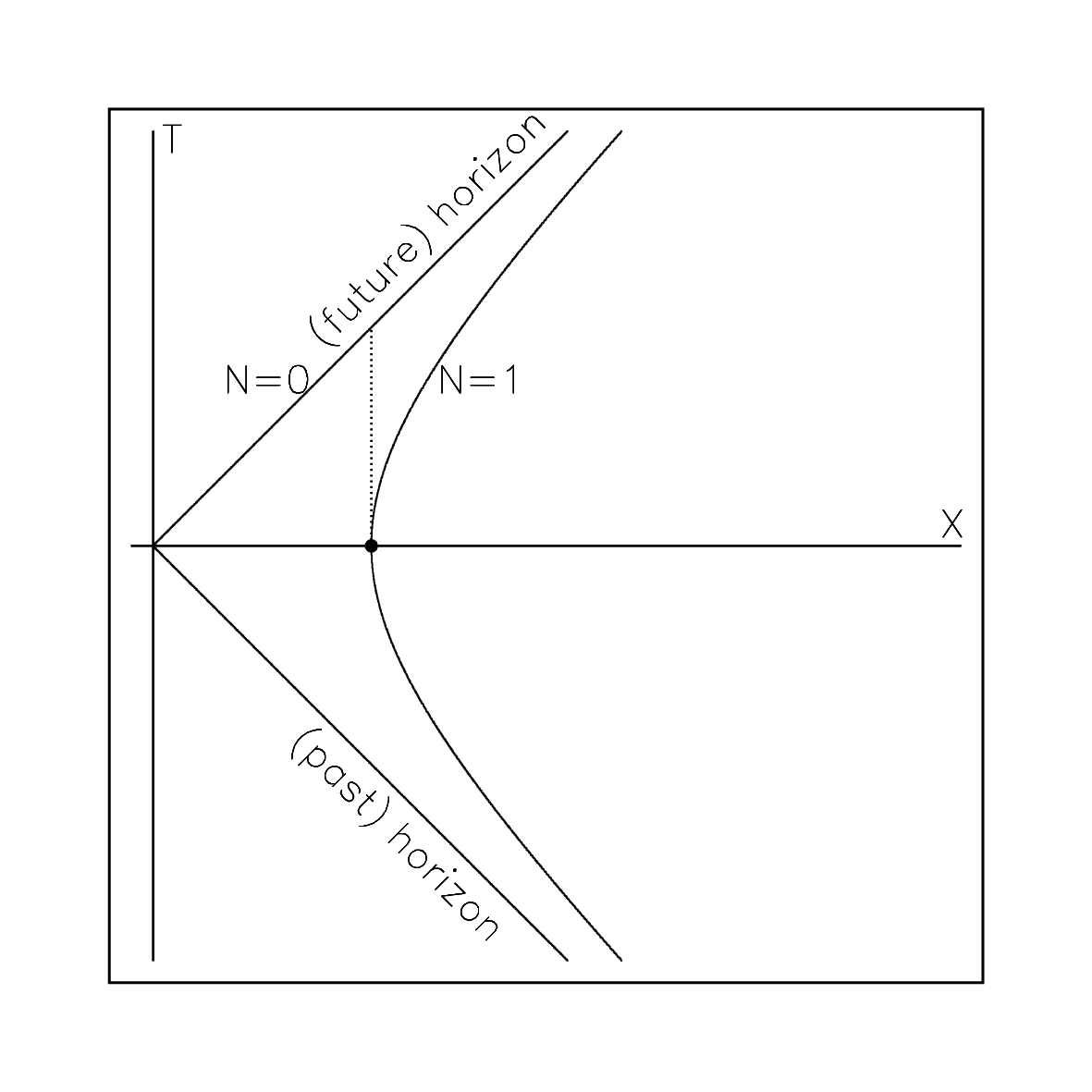}
\caption{(Piece of) worldline of an element of volume at $X = \frac{1}{\kappa}$,
at rest in the inertial frame (see text).}
\label{f}
\end{center}
\end{figure}

Let us now stay
with the element
of matter,
and consider any possible local Rindler horizon
with any direction and any temperature,
i.e. the horizon perceived by any possible
accelerating observer 
sitting
instantaneously 
where matter is, 
and about to absorb it.
We ask about
the relation between the entropy $dS$
of the lump of matter and the variation of 
the Clausius entropy of any of these horizons.
%
This corresponds 
to considering
the situation depicted
in Fig. \ref{f}, with $\{ X^a\}$ 
now being
the local frame
of matter and the element of matter located at $X = \frac{1}{\kappa}$
at $t = 0$,
and 
asking
about the relation between
$dS$ and $\frac{dE}{T_H}$.  

We evaluate $dS$
from Gibbs-Duhem relation
(for a 1-component thermodynamic system),
in its local form

\begin{eqnarray}\label{gib}
\rho =
s \theta - P + \mu n,
\end{eqnarray}
a relation
which merely expresses the first law of thermodynamics
joined to the request of (local) extensivity of 
pertinent thermodynamic potentials (cf. \cite{OppB}).
Here,
$\rho$, $s$, $\theta$, $P$, $\mu$ and $n$
are respectively local energy density, entropy, 
temperature,
pressure, chemical potential and number density
of the element of matter.
As we are concerned with gravity,
the rest-mass energy is thought here to be included in $\rho$;
we know this means that this same energy 
must be 
thought of as also included in $\mu$
(cf., for example, \cite{Hua} (p. 155)).
We have 

\begin{eqnarray}\label{dS}
dS =
s \ dV_{prop} =
\frac{\rho + P - \mu n}{\theta} \ dV_{prop},
\end{eqnarray}
where $dV_{prop}$ is the proper volume
of the element of matter. 

In the calculation of $\frac{dE}{T_H}$,
we explicitly choose
$T_{ab} = (\rho + P) u_a u_b +P g_{ab}$
($u^a$ is the velocity vector of the element of matter), 
that is we choose
the stress-energy tensor of an ideal fluid,
in the assumption that this case
is general enough 
to show what is 
here happening
concerning the relation
between $dS$ and $\frac{dE}{T_H}$, 
without, 
unnecessary at this stage,
further complications.
In (\ref{Clausius}),
all we need to know is 
$\xi^a_{(t)}$
on any given horizon.
From $\xi^a_{(t)} \rightarrow \kappa X k^a$
on the horizon  (cf., for instance, \cite{PadA}),
we get

\begin{eqnarray}\label{bound}
\frac{dE}{T_H} =
\frac{1}{T_H} \ T_{ab} \ k^a k^b \ dV_{prop},
\end{eqnarray}
and thus

\begin{eqnarray}\label{bound2}
\frac{dE}{T_H} =
\frac{\rho + P}{T_H} \ dV_{prop}.
\end{eqnarray}

Comparing $dS$ with $\frac{dE}{T_H}$, 
we see
amounts to compare
$s = \frac{\rho + P - \mu n}{\theta}$
with
$\frac{\rho + P}{T_H}$.
%
In this regard,
we notice that what we are talking about is an 
element of matter of some proper size $l$,
with temperature
and other thermodynamic potentials assigned,  
which, 
initially,
i.e.  at $t = 0$ and 
at a proper distance $D = \frac{1}{\kappa}$
from the horizon,
is assumed to be entirely ahead of the horizon.
This means 
$l \leq 2 D = \frac{2}{\kappa}$.

%
The circumstances 
can be conveniently described 
in terms of the length $l^*$ mentioned above,
characteristic of the given local thermodynamic conditions.
The meaning of $l^*$ is that,
for whatever thermodynamic system with given
(local) thermodynamic conditions, its size $\Delta$
cannot be smaller than the $l^*$ for these conditions.
%
More precisely,

\begin{eqnarray}\label{l*}
\Delta \geq
l^* \equiv
\frac{1}{\pi} \frac{s}{\rho + P} =
\frac{1}{\pi \theta} \Big( 1 - \frac{\mu n}{\rho + P} \Big),
\end{eqnarray}
where the 
last
equality follows from Gibbs-Duhem relation.
In practice this means that,
if we insist in saying that a system has certain
given values of $\rho$, $s$, and $P$,
we have to realize that that system cannot have size smaller
than $l^*$ as given in (\ref{l*}).

Inequality (\ref{l*}) 
is not universally proven,
it is a conjecture.
It is however likely to be obeyed
by any matter system, 
as an expression of the unavoidable quantum indeterminacy
of the constituent particles,
i.e. its reason rests on that
it seems to be simply a manifestation, or reformulation,
of the uncertainty principle.
As such, it can be considered as a universal feature
of any thermodynamic system,
and it is morover
a basic expression of quantum mechanics alone
(i.e. without
reference to gravity, meaning
with reference to Minkowski limit only).
In \cite{PesB, PesC}
an analysis is made on how
the inequality (\ref{l*}) arises
in some relevant and calculable systems.

Applying this here,
we get

\begin{eqnarray}
s \leq
\pi \ l (\rho + P) \leq
\frac{2 \pi}{\kappa} (\rho + P) =
\frac{\rho + P}{T_H},
\end{eqnarray}
and so

\begin{eqnarray}\label{end}
dS \leq
\frac{dE}{T_H},
\end{eqnarray}
where the equality is reached
when $l = l^*$ and $l = \frac{2}{\kappa}$.

The possibility of fulfilling
the condition $l = l^*$ 
is tied to
the nature of the matter under consideration.
For certain material media the equality   
can indeed in principle be reached;
for example
this happens for ultrarelativistic systems
with $\mu =0$, e.g. a blackbody photon gas, 
when the size $l$ of the system is made very small,
up to the
limit of compatibility, in view of quantum uncertainty, 
with the assigned values of thermodynamic potentials \cite{PesC}.
In general media, even when this quantum limit is reached,
$l$ is still much larger than $l^*$.
For generic choices of the size
of the element of matter, we will have $l \gg l^*$
even for photon gases.

The condition $l = \frac{2}{\kappa}$
deals with the location of the horizon
which 
absorbs
the element of matter.
When $l = \frac{2}{\kappa}$
the horizon just starts to absorb,
that is the element of matter is just ahead
of the horizon, at the limit of the beginning of the 
absorption;
and it is $l < \frac{2}{\kappa}$ otherwise,
i.e. when there is still some path before 
absorption
begins.
The eventuality $l > \frac{2}{\kappa}$ is not allowed,
since it is
incompatible with the assumption
that
the system is entirely ahead of the horizon at start;
a system with $l > \frac{2}{\kappa} = \frac{1}{\pi T_H}$,
where $T_H$ is the temperature 
of the horizon 
about to absorb it,
has necessarily a part of itself beyond the reach of the Rindler coordinates
of the accelerating observer located where matter is and
instantaneously at rest with respect
to it, whom
perceives that horizon,
so that this part of the system
results 
already 
absorbed
by the horizon.

We have thus shown the following.\\
{\it Proposition.}
Given an element of matter of size $l$ 
and energy-momentum tensor $T_{ab}$,
its entropy $dS$ is bounded from above
by the quantity $\frac{dE}{T_H}$, given by (\ref{bound}),
of a Rindler horizon
at temperature $T_H = \frac{1}{\pi l}$ as perceived 
by an accelerating observer
sitting instantaneously where matter is, 
with
$k^a$ the tangent vector to the generators 
of the horizon,
i.e. by the variation of Clausius entropy of a Rindler horizon
at the perceived temperature $T_H = \frac{1}{\pi l}$, 
which engulfs
the element of matter.

The limiting value can in principle be reached:
this happens for ultrarelativistic systems with $\mu =0$,
of very small size, pushed to the (quantum) limit of compatibility
with the assigned thermodynamic conditions 
for the element of matter.
The bound can 
thus also be considered
as the supremum
among the $dS$
for all possible material media, 
for given $dE$ and $l$.
Thus considered,
it is manifest that the bound is a quantity inherent to matter,
and 
is
irrespective of the gravity theory,
in agreement with what we already know
since, as seen, the quantity $\frac{dE}{T_H}$ is independent
of the gravity theory.

\section{The generalized covariant entropy bound extended 
to general theories of gravity}

In the descriptions-derivations of the equations
of motion as a thermodynamic identity,
the equality

\begin{eqnarray}{\label{dSH}}
\frac{dE}{T_H} = dS_W
\end{eqnarray}
between the
Clausius entropy variation of the horizon to the left, 
and the Wald 
entropy variation
of horizon to the right
(coinciding with the Bekenstein-Hawking 
entropy variation $dS_{BH}$
in Einstein theory) 
associated with
the crossing of the horizon
by some element of matter 
is stressed as entailing
the equations of motion for the gravitational field, 
i.e. how matter curves geometry.
Indeed,
Wald's entropy $S_W$
is a prescription which, 
on using
the equations of motion, 
gives the amount of entropy 
associated with
a patch of horizon of some $(D-2)$-dimensional cross-sectional area $A$,
the amount of entropy 
being
different for different gravity theories. 
%
The l.h.s.
is clearly independent of the gravity theory,
as can be 
seen,
for example, 
on noting
that any bifurcate horizon can be approximated
at a point by a Rindler horizon, and applying
the discussion of previous Section.
The r.h.s. must also be independent
of the gravity theory.
%
Equality (\ref{dSH}) 
then gives
how much the area $A$ must shrink,
due to some passage of matter,
this depending
on the expression of $S_W$.
For each expression of $S_W$
this gives the equations of motion for the theory
which has that expression as its Wald's entropy of the horizon. 

Using this,
the bound discussed in the previous Section
becomes, in each specific theory of gravity,
a bound in terms of the Wald entropy of that theory.
From (\ref{end}) and (\ref{dSH}), in fact,
we get

\begin{eqnarray}\label{gen}
dS \leq
dS_W.
\end{eqnarray}

According to previous Section
we are considering
this relation in terms of
a patch of Rindler horizon at $P$ 
with tangent $k^a$ to the generators 
and temperature $T_H = \frac{\kappa}{2 \pi}$.
We can however think of $dS_W$
as referring to a generic bifurcate Killing horizon $H$ at $P$
with normal $\xi_{H}$,
tangent $k_H^a$ to the generators
and surface gravity $\kappa_H$,
on calculating it
through the approximating
Rindler patch at $P$ with $\xi_{(t)} = \xi_{H}$,
$k^a = k_H^a$
and $\kappa = \kappa_H$.

In $D-$dimensional Einstein gravity
$S_W$ coincides with $S_{BH}$,
and (\ref{gen}) becomes


\begin{eqnarray}
dS \leq
dS_{BH} =
\frac{dA}{4}.
\end{eqnarray}
%
But, this is 
the generalized covariant entropy bound \cite{Fla} (GCEB),
as applied to the patch of $(D-2)$-hypersurface of area $A$ 
coinciding with the $(D-2)$-section of 
a patch of a horizon
and having $k_H^a$ as the null field orthogonal to it.
%
Inequality (\ref{gen}) 
thus generalizes
the GCEB to a form which applies in any metric theory of gravity,
with the generalization being in that
$\frac{dA}{4}$,
with $A$ the $(D-2)$-area of a patch $B$
of $(D-2)$-hypersurface 
with an orthogonal
null field $k^a$, 
is replaced by $dS_W$,
thought to be
the Wald entropy
of the patch 
of the sheet of null hypersurface
generated by null geodesics which start
at $B$ with tangent $k^a$.

\section{$l^*$ for horizons}

The scale $l^*$ has been considered in the past
for conventional matter only \cite{PesB, PesC}.
The discussion of Section \ref{section2}
gives the opportunity 
of defining
a value $l^*_H$ for it
for horizons.  
The definition of $l^*$ given in (\ref{l*})
can evidently be put also in the form

\begin{eqnarray}\label{l*_2}
l^* \equiv
\frac{1}{\pi} \frac{s \ dV_{prop}}{(\rho + P) dV_{prop}} =
\frac{1}{\pi} \frac{dS}{dE}.
\end{eqnarray}
This is the $l^*$ of the matter 
which is going to be 
swallowed by a horizon.
The matter has 
an entropy content $dS$
in some proper volume $dV_{prop} = A l$
and gives an energy contribution $dE$ to the horizon,
which has 
a given temperature $T_H$.
%
Now,
if we use matter for which the limit
$l = l^*$ can be reached, 
and we choose $l = l^*$,
and if we furthermore
assume that our choice of the thermodynamic parameters of matter
is such that 
$l = l^*$ is the maximum value allowed for the matter to
be absorbed in one bite by the given horizon,
i.e. $l = \frac{1}{\pi T_H}$, 
we know that in (\ref{end}) the equality holds, 
and we are thus 
allowed
to write 
$\frac{dE}{T_H}$ for $dS$ in (\ref{l*_2}). 
But, 
the member to the right in (\ref{l*_2}) 
is
now given
in terms of quantities which refer to the 
to the horizon alone.
This suggests

\begin{eqnarray}\label{l*H}
l^*_H \equiv
\frac{1}{\pi} \frac{dE/T_H}{dE} =
\frac{1}{\pi T_H} =
\frac{1}{\pi} \Big( \frac{s}{\rho + P} \Big)_{l = l^* = \frac{1}{\pi T_H}},
\end{eqnarray}
where the subscript reminds us that
the term 
in the round brackets
must be evaluated
for material media for which 
the choice
$l = l^* = \frac{1}{\pi T_H}$
is allowed.
We said above that
a blackbody photon gas,
for which $l^* = \frac{1}{\pi \theta}$ (cf. (\ref{l*})),
does the job in
that it allows for $l = l^*$
and we get  $l = \frac{1}{\pi T_H}$
provided the temperature of the gas 
equals 
the horizon
temperature, $\theta = T_H$.  
$l^*_H$ is 
thus the order
of the wavelength
of blackbody photons  
at the temperature $T_H$ of the thermal bath perceived
by 
the accelerating observer.
%
This quantity is supposed to conveniently characterise
the behaviour of a horizon from a pure quantum mechanical
point of view, meaning in the Minkowski limit
(and we know horizons do not imply we leave this limit).
As such, it does not contain any information on gravity,
and is well defined also in a context 
for which,
spacetime 
is purely Minkowskian (cf. \cite{PesC}).
It is therefore
a concept clearly distinct from the Planck scale $l_P$
(and, in general, is enormously larger than the latter).

For a generic bifurcate Killing horizon,
its $l^*_H$ can be defined in an obvious way through
the Rindler horizon approximating it at a point.
For systems collapsing to form
black holes in Einstein gravity with temperature $T_H$
in asymtotically-flat spacetimes, for example,
the photon gas (or ultrarelativistic matter) above, 
which gives $l = l^* = \frac{1}{T_H}$,
is a blackbody
and must have temperature $\theta = T_H$,
as measured by 
distant observers
(as can be verified 
from locally Rindler approximating the horizon);
i.e., it turns out 
it is just their Hawking radiation.
For the Schwarzschild black hole
we get

\begin{eqnarray}
l^*_H =
\frac{1}{\pi T_H} =
\frac{8 \pi M}{\pi} =
8 M =
4 R,
\end{eqnarray}
where $M$ and $R$ are black hole mass and radius.

\section{Concluding remarks}

We have seen
the
limiting 
thermodynamical scale
$l^*$ of matter
can be used to show 
that entropy of any element of matter 
is bounded by the variation of Clausius entropy 
of a suitable Rindler horizon 
absorbing it.
%
From this, a form of the GCEB, 
valid for general metric theories of gravity,
has been introduced,
and the $l^*$ for horizons has been defined.

Using the perspective according to which
the entropy of any patch of 
horizon 
is always (i.e. for any diff-invariant theory of gravity)
given by a quarter of area
in units of effective coupling \cite{BruB},
from (\ref{gen}) we get

\begin{eqnarray}\label{bru}
dS \leq
\frac{dA}{4 G_{eff}},
\end{eqnarray}
where the effective coupling $G_{eff}$,
defined as in \cite{BruB},
is explicitly
reported,
and, again, $A$ is the area of a patch of the boundary surface.
$G_{eff}$ 
depends in general on the point in spacetime,
and $G_{eff}$ equals 
Newton's constant,
$G_{eff} = $ constant
$ = G_N$, for Einstein gravity. 
Formula (\ref{bru}) is a further (local) expression 
of the GCEB generalized to
general theories of gravity.
When we change the gravity theory,
the quantity $\frac{dE}{T_H}$ and thus $\frac{dA}{4 G_{eff}}$
remains unaffected.
%
What changes,
are $dA$ and $G_{eff}$ individually.
A gravity theory with a gravitational coupling, say, 
larger than that of Einstein's theory ($G_{eff} > G_N$), 
will give
a stronger focusing (a larger $dA$),
stronger by the amount exactly needed to leave
 $\frac{dA}{4 G_{eff}}$ invariant.

The fact that
the black hole horizon has the same $l^*$ as
its Hawking radiation,
can be interpreted
as suggesting that
in some respects 
it behaves like the Hawking radiation itself.
%
This radiation is, after all, all that
a distant observer can see of the black hole.
This thermodynamic fact, could provide some further understanding
of the dynamic
behaviour of black holes.
The 
dynamic relaxation times
of perturbed black holes (for a review see \cite{Nol}),
could be seen
as an example of this kind.
The argument for this, goes as follows.
A universal lower limit 
$\tau_0(\theta) \equiv
\frac{1}{\pi \theta}$
to the relaxation times
$\tau$ of systems at temperature $\theta$, 
has been introduced 
in \cite{HodA}
from thermodynamics and quantum information theory.
As far as black holes can be considered thermodynamic objects,  
we would expect that $\tau_0(T_H)$ 
 could also
limit the dynamic relaxation times of perturbed black holes
with horizon temperature $T_H$.
Could this be inspected through 
the use
of $l^*$ for horizons?
Systems consisting of blackbody radiation
at temperature $\theta$
actually obey this limit,
and,
shaped as thin layers,
they exhibit, 
when sufficiently thin, 
relaxation times $\tau$ approaching it,
$\tau = \tau_0(\theta)$ 
(in the limit $l \rightarrow l^*(\theta)$,
$l$ being their thickness,
limit that, as mentioned above, for blackbody radiation can be reached)
\cite{PesD}.
Now,
the imprint on any radiation, 
of the variation of the properties of the black hole
cannot have any evolution more rapid
than 
that given
by this limit with $\theta = T_H$.
That means,
the distant observer
cannot see
variations of the black hole
on a characteristic scale smaller than this,
i.e. the relaxation times of black holes
cannot be smaller than this limit:
$\tau \geq \tau_0 = \frac{1}{\pi T_H}$.
This, 
namely that black holes comply with the bound,
is precisely what is obtained
by analytical and numerical evaluations
\cite{HodA}.
In \cite{HodA} it is found moreover that,
when black holes are nearly extremal, 
i.e. in the $T_H \rightarrow 0$ limit,
they saturate the bound.
According to the perspective above, 
near-extremal black holes could thus be regarded
as equivalent
to a thickness $l = l^*_H = \frac{1}{\pi T_H}$ of
their own Hawking radiation,
from a far observer point of view. 

A last comment concerns the particular 
case of matter for which $P = -\rho$,
i.e. matter mimicking the effects of
a cosmological constant term in the equations of motion.  
For this matter,
the Gibbs-Duhem relation
(\ref{gib}) becomes

\begin{eqnarray}\label{gib_2}
s \theta + \mu n = 0,
\end{eqnarray}
where $s$, with its informational meaning,
and $\theta$ and $n$ should be regarded
as intrinsically non-negative.
We have two possibilities:\\
i) $s > 0$; implying $\mu < 0$;\\
ii) $s = 0$; implying $\mu n = 0$.\\
Looking at (\ref{bound2}),
both cases give $\frac{dE}{T_H} = 0$,
and we do not get an increment
of Clausius entropy of the horizon
when such matter crosses it.
In case (i),
this gives a violation of (\ref{end}), and thus of (\ref{gen}).
In $l^*$ terminology, 
the situation is characterized by $l^* = \infty$ in (\ref{l*}),
and thus 
we see that
we can never have $l \ge l^*$ for this matter for any $l$;
that is,
matter of this kind is incompatible
with the assumption 
for it
to be initially
completely ahead of 
the horizon.
%
The basic assumption itself that matter can be described in terms
of local quantities ($\rho$, $s$, ...)
in a volume $dV_{prop}$,
a pre-requisite for writing the local form of 
the Gibbs-Duhem relation
(\ref{gib}), 
is put into question;
the condition itself $s > 0$
becomes of doubtful meaning
in that $s$ 
is found to be
ill-defined.
In view of this,
we cannot claim that any matter of this kind
would produce a violation of entropy bound (\ref{gen});
what we could claim
is that the bound could hold true even in presence
of this ``matter'',
since the real point is that 
surely
our description of it in local terms
is no longer adequate
(this perhaps offers a different perspective
on the issue of the effect of any cosmological constant term
on entropy bounds as tackled in \cite{LeeB}).   
We could speculate that,
if we believe in quantum mechanics as providing $l^*$,
the case $s > 0$ would correspond
to ``cosmological matter'' which would 
intrinsecally be
completely delocalised
($l^* = \infty$).
If we think of constituents for it, 
they should be completely delocalised,
and maximally entangled.

Case (ii) gives
$l^* = \frac{0}{0}$,
that is $l^*$ is undetermined without further input.
Any finite value of $l^*$ would allow inequality (\ref{end})
to be satisfied
irrespective of $l$,
i.e. even when $l < l^*$.
This would suggest, for overall consistency, $l^* = 0$ for case (ii).   
It would correspond to matter completely localised,
with no entanglement, 
with infinite energy and pressure (from quantum indeterminacy),
and no entropy.

{\it Acknowledgments.}
%
I thank
Giovanni Venturi
for the careful reading of the manuscript.


\begin{thebibliography}{00}

\bibitem{JacA}
T. Jacobson,
``Thermodynamics of spacetime: the Einstein equation of state'',
Phys. Rev. Lett. {\bf 75} (1995) 1260, 
gr-qc/9504004.

\bibitem{PadE}
T. Padmanabhan, A. Paranjape,
``Entropy of null surfaces and dynamics of spacetime'',
Phys. Rev. D {\bf 75} (2007) 064004,
gr-qc/0701003.

\bibitem{PadF}
T. Padmanabhan,
``Dark energy and gravity'',
Gen. Rel. Grav. {\bf 40} (2008) 529,
arXiv:0705.2533.

\bibitem{PadA}
T. Padmanabhan,
{\it Gravitation: Foundations and frontiers}
(Cambridge Univ. Pr., Cambridge UK, 2010).

\bibitem{PadB}
T. Padmanabhan,
``Thermodynamical aspects of gravity: new insights'',
Rept. Prog. Phys. 73 (2010) 046901, 
arXiv:0911.5004.

\bibitem{JacC}
T. Jacobson, R. Parentani,
``Horizon entropy'',
Found. Phys. {\bf 33} (2003) 323,
gr-qc/0302099.

\bibitem{Fla}
$\acute {\rm E}$.$\acute {\rm E}$. 
Flanagan, D. Marolf and R.M. Wald,
``Proof of classical versions of the Bousso entropy bound and of the 
generalized second law'',
Phys. Rev. D {\bf 62} (2000) 084035, 
hep-th/9908070.

\bibitem{BekD}
J.D. Bekenstein,
``Universal upper bound on the entropy-to-energy ratio  
for bounded systems'',
Phys. Rev. D {\bf 23} (1981) 287.

\bibitem{JacB}
T. Jacobson,
``Gravitation and vacuum entanglement entropy'',
arXiv:1204.6349 (2012).

\bibitem{Ver}
E.P. Verlinde,
``On the origin of gravity and the laws of Newton'',
JHEP {\bf 1104} (2011) 029,
arXiv:1001.0785.

\bibitem{PesG}
A. Pesci, 
``Gravity from the entropy of light'',
Class. Quantum Grav. {\bf 28} (2011) 045001, 
arXiv:1002.1257.

\bibitem{PesH}
A. Pesci 
``The existence of a minimum wavelength for photons'',
arXiv:1108.5066 (2011).

\bibitem{PadD}
T. Padmanabhan,
``Entropy of static spacetimes and microscopic density of states'',
Class. Quantum Grav. {\bf 21} (2004) 4485,
gr-qc/0308070.

\bibitem{PadC}
T. Padmanabhan,
``Equipartition of energy in the horizon degrees of freedom and the emergence 
of gravity'',
Mod. Phys. Lett. A 25 (2010) 1129,
arXiv:0912.3165.

\bibitem{Cai}
R.-G. Cai, S.P. Kim,
``First law of thermodynamics and Friedmann equations of 
Friedmann-Robertson-Walker universe'',
JHEP {\bf 0502} (2005) 050,
hep-th/0501055.

\bibitem{LanA}
C. Lanczos,
``Electricity as a natural property of Riemannian geometry'',
Rev. Mod. Phys. {\bf 39} (1932) 716.

\bibitem{LanB} 	
C. Lanczos,
``A remarkable property of the Riemann-Christoffel tensor in four dimensions'',
Annals Math. {\bf 39} (1938) 842.

\bibitem{Lov}
D. Lovelock,
``The Einstein tensor and its generalizations'',
J. Math. Phys. {\bf 12} (1971) 498.

\bibitem{Eli}
E. Elizalde and P.J. Silva, 
``F(R) gravity equation of state'', 
Phys. Rev. D {\bf 78} (2008) 061501, 
arXiv:0804.3721.

\bibitem{BruC}
R. Brustein, M. Hadad,
``The Einstein equations for generalized theories of gravity 
and the thermodynamic relation $\delta Q = T \delta S$ are equivalent'',
Phys. Rev. Lett. {\bf 103} (2009) 101301, 
Erratum-ibid. {\bf 105} (2010) 239902,
arXiv:0903.0823.

\bibitem{WalB}
R.M. Wald,
``Black hole entropy is Noether charge'',
Phys. Rev. D {\bf 48} (1993) 3427,
gr-qc/9307038.

\bibitem{JacD}
T. Jacobson, G. Kang, R.C. Myers,
``On black hole entropy'',
Phys.Rev. D {\bf 49} (1994) 6587,
gr-qc/9312023.

\bibitem{Iye}
V. Iyer and R.M. Wald,
``A comparison of Noether charge and Euclidean methods for computing the 
entropy of stationary black holes'',
Phys. Rev. D {\bf 52} (1995) 4430,
gr-qc/9503052.

\bibitem{KotC}
D. Kothawala,
``The thermodynamic structure of Einstein tensor'',
Phys. Rev. D {\bf 83} (2011) 024026,
arXiv:1010.2207.

\bibitem{Bac}
V. Baccetti, M. Visser,
``Clausius entropy for arbitrary bifurcate null surfaces'',
arXiv:1303.3185.

\bibitem{PesB}
A. Pesci,
``From Unruh temperature to the generalized Bousso bound'',
Class. Quantum Grav. {\bf 24} (2007) 6219, 
arXiv:0708.3729.

\bibitem{PesC}
A. Pesci,
``On the statistical-mechanical meaning of the Bousso bound'',
Class. Quantum Grav. {\bf 25} (2008) 125005,
arXiv:0803.2642.

\bibitem{RovC}
C. Rovelli,
{\it Quantum gravity}
(Cambridge Univ. Pr., Cambridge UK, 2004).

\bibitem{OppB}
J. Oppenheim,
``Thermodynamics with long-range interactions: from Ising models
to black-holes'', 
Phys. Rev. E {\bf 68} (2003) 016108,
gr-qc/0212066.

\bibitem{Hua}
K. Huang,
{\it Statistical Mechanics} (J. Wiley \& Sons, New York, 1987).

\bibitem{BruB}
R. Brustein, D. Gorbonos, M. Hadad,
``Wald's entropy is equal to a quarter of the horizon area 
in units of the effective gravitational coupling'',
Phys. Rev. D {\bf 79} (2009) 044025,
arXiv:0712.3206.

\bibitem{Nol}
H.-P. Nollert,
``Quasinormal modes: the characteristic `sound' of black holes and 
neutron stars'',
Class. Quantum Grav. {\bf 16} (1999) R159.

\bibitem{HodA}
S. Hod,
``Universal bound on dynamical relaxation times and black-hole quasinormal  
ringing'',
Phys. Rev. D {\bf 75} (2007) 064013,
gr-qc/0611004.

\bibitem{PesD}
A. Pesci, 
``A note on the connection between the universal relaxation bound and the 
covariant entropy bound'',
Int. J. Mod. Phys. D {\bf 18} (2009) 831,
arXiv:0807.0300.

\bibitem{LeeB}
J.-W. Lee,
``Zero Cosmological Constant and Nonzero Dark Energy from Holographic 
Principle'',
J. Korean Phys. Soc. {\bf 63} (2013) 1088,
arXiv:1003.1878.

 	



\end{thebibliography}
\end{document}